\documentclass[letterpaper, 10 pt, conference]{ieeeconf}  

\IEEEoverridecommandlockouts                              

\usepackage{cite}
\usepackage{graphicx}
\usepackage{amsmath} 
\usepackage{hyperref}
\usepackage{float}
\usepackage{amsmath}
\usepackage{multirow}
\usepackage{booktabs}
\usepackage{makecell}
\usepackage{siunitx}
\usepackage{stfloats}
\usepackage{orcidlink}
\usepackage{xcolor} 

\title{\LARGE \bf
Respiratory Motion Compensation and Haptic Feedback for X-ray-Guided Teleoperated Robotic Needle Insertion
}

\author{Ana Cordón-Avila\orcidlink{0009-0003-2148-5457}, Mostafa Selim\orcidlink{0000-0002-5224-4790} and Momen Abayazid\orcidlink{0000-0003-1857-5921}
\thanks{*This work is part of the 20044 ASSIST project and funded by ITEA}
\thanks{All authors are affiliated with the Robotics and Mechatronics research group, TechMed Centre, Faculty of Electrical Engineering, Mathematics and Computer Science, University of Twente, 7500 AE Enschede, The Netherlands.
Email:~\href{mailto:a.cordon@utwente.nl}{a.cordon@utwente.nl};~\href{mailto:m.s.selim@utwente.nl}{m.s.selim@utwente.nl}; 
~\href{mailto:m.abayazid@utwente.nl}{m.abayazid@utwente.nl}}%
}

\begin{document}

\maketitle
\thispagestyle{empty}
\pagestyle{empty}

\begin{abstract}
Respiratory motion limits the accuracy and precision of abdominal percutaneous procedures. 
In this paper, respiratory motion is compensated robotically using motion estimation models. 
Additionally, a teleoperated insertion is performed using proximity-based haptic feedback to guide physicians during insertion, enabling a radiation-free remote insertion for the end-user. 
The study has been validated using a robotic liver phantom, and five insertions were performed. 
The resulting motion estimation errors were below 3~mm for all directions of motion, and the overall resulting 3D insertion errors were 2.60~mm, 7.75~mm, and 2.86~mm for the superior-inferior, lateral, and anterior-posterior directions of motion, respectively. 
The proposed approach is expected to minimize the chances of inaccurate treatment or diagnosis due to respiratory-induced motion and reduce radiation exposure. A video summary can be found in: \href{https://youtu.be/2Xw6AMe2mco}{https://youtu.be/2Xw6AMe2mco} 

\end{abstract}
\section{INTRODUCTION}
The Global Cancer Observatory reported in 2020 that the incidence of liver cancer accounted for approximately 905,000 cases worldwide \cite{c1}. 
Percutaneous liver biopsy is considered the gold-standard technique to diagnose hepatic lesions \cite{c2}. 
Ultrasound (US) or computed-tomography (CT) scans are usually used to guide physicians throughout the insertion \cite{c3}. 
US provides real-time guidance without ionizing radiation \cite{c4}. 
However, there is high interoperator variability and the spatial resolution is limited~\cite{c5}. 
CT-guidance can be preferred when there is difficulty in visualizing lesions under US \cite{c4,c6}. 
Some limiting factors of X-ray-guided percutaneous procedures are radiation exposure, increase in interventional time, and increase in costs \cite{c6}.

The precision of needle placement in percutaneous procedures is significantly affected by respiratory motion~\cite{c5}.  
Physicians advance the needle at the end-expiration phase or during a breath-hold~\cite{c5,c8}.
However, some patients might not be able to hold their breath for sufficient time. 
For those interventions in which the needle is not completely inserted within the same breath-hold, respiratory-induced motion can lead to inaccurate needle placement~\cite{c5}. 
In such cases, respiratory motion management leads to several extra scans to ensure that the needle advances along the planned path~\cite{c8}. 
This leads to an increase in radiation exposure and interventional time~\cite{c7}. 
Alternatively, real-time respiratory motion models offer the possibility to estimate and compensate for respiratory motion. 

Real-time respiratory motion models based on external signals offer the opportunity to estimate the location of the tumor and thus compensate for respiratory-induced motion. 
These external signals are known as surrogate signals and their objective is to capture the external changes due to breathing to estimate the tumor position. 
These signals need to be easy to measure, be highly correlated with the induced motion at the region of interest, and have high-frame rate~\cite{c10}. 
The correlation between the surrogate signal and the tumor displacement can be developed prior to the intervention and the created motion model can be tested during the procedure~\cite{c10}. 
Motion models can provide the precision needed in hepatic percutaneous procedures while decreasing the number of CT images required for targeting the lesion since the real-time position of the tumor is known~\cite{c11}. 
Additionally, it is not required for the patient to maintain a specific breath-hold position throughout the intervention since motion models are able to estimate the location of the tumor at any breath-hold position~\cite{c9}.

Needle alignment can be achieved by integrating motion models and robotics.
The model can update the pose of a robotic arm to steer the needle towards the moving target and thus compensate for the respiratory-induced motion.
Several studies have implemented robotic manipulators in X-ray-based image-guided percutaneous procedures~\cite{c7,c12,c13, c14, c15}. 
Robotic arms offer the opportunity of increasing stability, accuracy, and dexterity during needle positioning with fewer adjustments. 
Accurate placement is important to avoid loss of time, additional X-ray exposure, misdiagnosis in liver biopsy, or incomplete treatment in tumor ablation~\cite{c16}. 
In~\cite{c12}, the Epione (Quantum Surgical, Montpellier, France) system was used in a clinical study to perform robotic-assisted CT-guided percutaneous needle insertion for hepatic thermal ablation. 
Other attempts for robotic CT-guided biopsies have shown that reducing the number of check scans can reduce the duration of the procedure and the radiation dose without sacrificing the accuracy of the procedure~\cite{c7}.

Additionally, robotic arms can be teleoperated to perform surgical tasks such as needle insertion. A teleoperated X-ray-guided insertion would enable the interventional radiologist to perform the procedure without radiation exposure. 
However, the loss of needle depth perception relative to the targeted tumor might affect the insertion accuracy.
Haptic feedback has been shown to enhance proximity perception to target objects and for spatial navigation~\cite{c17,c18,c19}.
Mieling et al. proposed a proximity-based haptic feedback method for epidural anesthesia\cite{c20}. 
The method reduced unintentional punctures and improved needle placement compared to visual and force-based haptic feedback.
Moreover, Dagnino et al. developed an endovascular catheterization teleoperation system with vision-based haptic feedback that reduced vessel-collision forces~\cite{c21}. 
Haptic cues were provided to be aware of the proximal vessels to the catheter through viscous forces and to detect collision incidents.
Using haptic feedback during needle insertion provides reliable real-time guidance to surgeons, potentially mitigating radiation exposure to the patients by reducing the number of extra scans. 

The contribution of this work includes the integration of respiratory motion compensation with teleoperated needle insertion for robotic X-ray-guided percutaneous procedures. 
Respiratory motion models estimate the position of a target in the liver, and the motion is compensated by robotic needle steering.
The remote insertion is performed with target proximity-based haptic feedback. 
Additionally, physicians perform a teleoperated insertion, eliminating radiation exposure during the procedure. 
Aspects other than respiratory motion that influence the accuracy of needle insertion, such as needle deflection during insertion and tissue deformation due to tool-tissue interaction are beyond the scope of this study. 
The proposed system aims to enhance physicians' guidance during insertion using haptic feedback while increasing the accuracy of the procedure using robotic motion compensation.

\begin{figure*}\label{fig:Fig1_sysarch}
    \vspace{5mm}
    \centering    
    \includegraphics[scale=0.7]{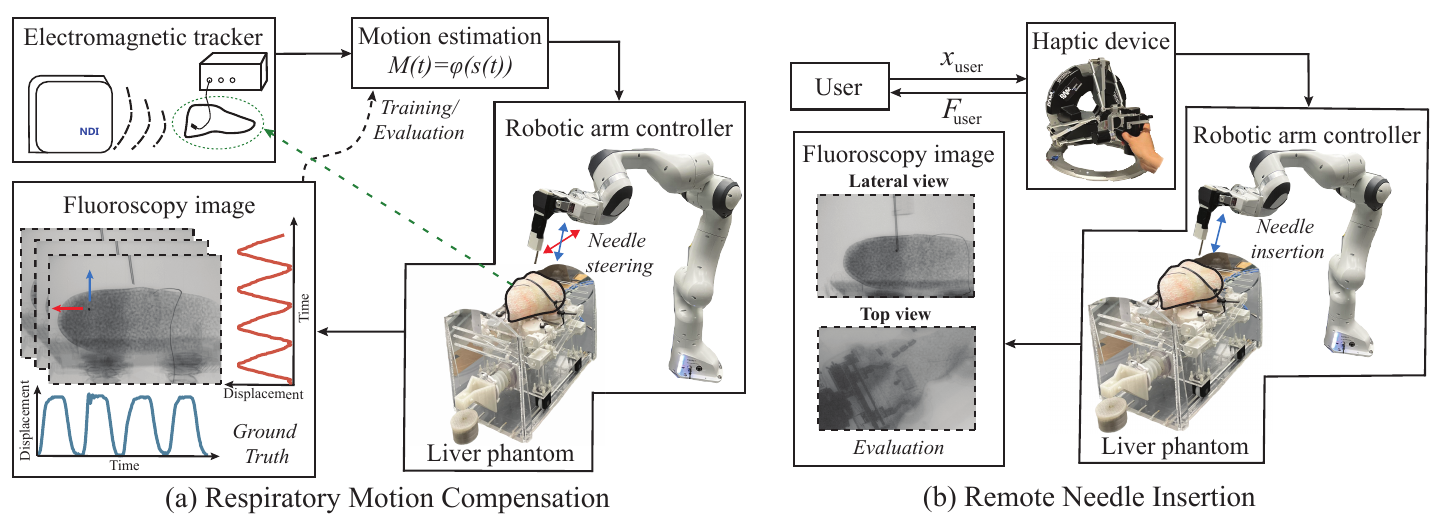}
    \caption{(a) The respiratory motion compensation step uses an electromagnetic tracker attached to the liver phantom to create a motion model. The model is trained using as ground truth the motion of the target inside the liver phantom extracted from fluoroscopic images. The robotic arm with a needle at its end-effector compensates for the respiratory motion using the output of the motion model. The robot performs needle steering to maintain the alignment between the target and the needle. (b) In the remote needle insertion step, the end-user remotely controls the robotic arm by operating the haptic device with force feedback ($F_{\text{user}}$), guiding the physicians to reach the desired displacement ($\textit{x}_{\text{user}}$).}
\end{figure*}
\section{MATERIALS AND METHODS}
The proposed approach consists of two main steps; firstly, the respiratory motion compensation, and secondly, the remote needle insertion. 
The motion compensation step uses an electromagnetic (EM) tracker as a surrogate signal to estimate the target motion. 
The EM sensor is attached externally to a liver phantom.  
The phantom simulates the respiratory-induced motion in the liver in the Anterior-Posterior (AP) and Superior-Inferior (SI) directions of motion. 
A spherical lead target visible under X-rays is placed inside the liver phantom to simulate the tumor.  
A motion model is trained using the EM sensor data as a surrogate, and the actual target motion as ground truth. 
The target displacement is extracted using fluoroscopy images. 
The robotic manipulator uses the estimated target location to compensate for the respiratory motion by steering a needle attached to its end-effector.
Needle steering preserves the alignment between the needle and the target.
The second step involves the remote teleoperation of the robotic arm. 
An end-user uses a haptic device to control the motion of the arm to perform the needle insertion task~(Fig.~1). 

\subsection{Respiratory Motion Compensation}

\subsubsection{Surrogate Signal} 
An EM sensor was implemented to track the displacement of the liver phantom. 
The detected displacements in the y- and z-axes of the sensor served as a surrogate signal. 
In~\cite{c24} and~\cite{c25}, it was observed that EM trackers can be used in human subjects as surrogate signals when attached to the abdominal area. 
In this work, the sensor was attached externally to the liver phantom. 
The obtained motion simulates the external abdominal skin motion during respiration. 

\subsubsection{Ground Truth} 
The ground truth represents the actual motion of the target. 
It is mainly required in the training phase of motion models and is used for validation purposes during the test phase.
For the present experiments, an angiography scan was used to extract the SI and AP respiratory-induced motion of the target. 
The training phase involves acquiring lateral fluoroscopic images of the liver phantom for 30~seconds at 15 frames per second. 
The motion of the liver phantom during the training time consisted of 12~seconds of regular breathing, and three 5-second breath-holds at three different phases of the respiratory cycle (target positions). 
Simultaneously, the external motion tracked by the EM tracker is measured and used as input for the motion model. The EM data and target motion are synchronized by introducing changes in the respiratory phase at the beginning and end of the training. 

\subsubsection{Correspondence Model} 
Two motion models were created for each breathing phase to estimate the respiratory-induced motion in the SI and AP directions. 
Eq.~\ref{simple_relation} depicts the general relationship that defines a motion model. The target location or ground truth signal ($M(t)$)
is estimated fitting the surrogate signal ($s(t)$) into the function $\psi$.
\begin{equation}
{M(t)= \psi (s(t))}
\label{simple_relation}
\end{equation}
A polynomial regression model was selected to define the correspondence between the target motion and the EM sensor data. 
\begin{equation}
{y_i \approx y_i(\beta) = \beta_0 + \beta_1x + \beta_2x^2+ ... +\beta_nx^n}
\label{eq1_regr}
\end{equation}
The \begin{math} {\beta} \end{math} coefficients of the model are determined during the training phase using the ordinary least squares method. 
Each model was validated by computing the mean absolute error (MAE) between the ground truth (\begin{math} { y_i} \end{math}) and the estimation (\begin{math} { y_i(\beta)} \end{math}) obtained using the surrogate signal (\begin{math} {x} \end{math}).

\begin{figure*}[!t]
  \centering
    \vspace{5mm}
\includegraphics{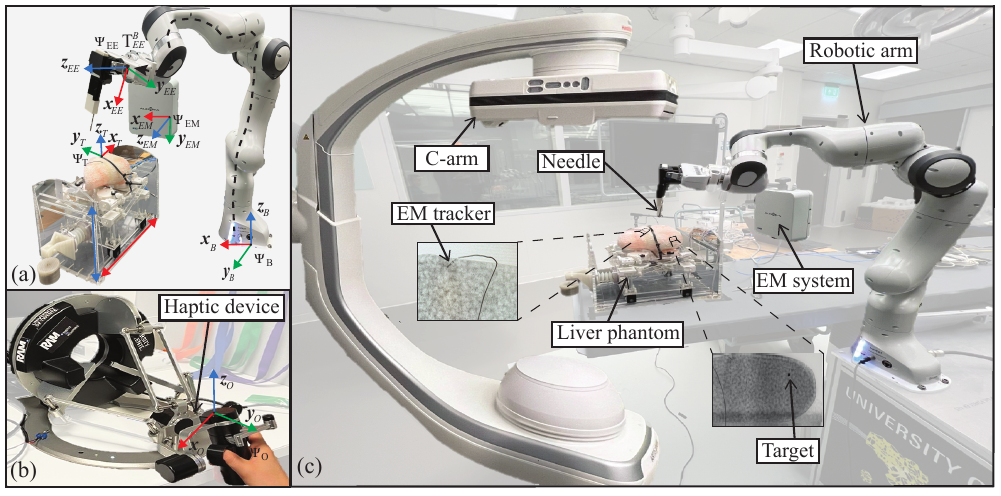}
  \caption{(a) A transformation matrix from the base frame ($\Psi_B$) to the end-effector  ($\Psi_{EE}$) realigns the robotic arm with the frame of the moving target ($\Psi_T$). The arm moves based on the estimations made using the electromagnetic (EM) system as a surrogate. The estimations relate the real-time position of the sensor in the EM frame ($\Psi_{EM}$) to the target frame ($\Psi_T$). (b) The source side contains the haptic device controlled by the user to move the robotic arm. The user moves the handle in the $y_o$ direction relative to the initial location of frame ($\Psi_O$). (c) The replica side contains the liver phantom with a spherical lead target located inside the phantom. The EM tracker system is attached to the phantom and used as a surrogate signal, and the C-arm scan is used to train and validate our methods. The robotic arm with a needle at the end-effector enables respiratory motion compensation and remote insertion steps.}
  \label{fig:setup}
  \vspace{-3mm}
\end{figure*}

\subsubsection{Robotic Motion Compensation} 
Motion models estimate the target location in the SI and AP directions of motion and the estimated displacements enable the adjustment of the robotic manipulator end-effector. 
This motion aligns the needle with the target correcting for the respiratory-induced motion and enabling a pre-insertion autonomous alignment.

\subsection{Teleoperated Needle Insertion with Haptic Feedback}
The end-user operates the haptic device to remotely control the robotic arm.
The displacement of the handle of the haptic device is directly mapped to the displacement in the axial direction of the needle. 
The displacement needed to reach the target is defined based on an initial cone beam CT 3-D reconstruction of the liver phantom. A manual registration between the needle and the liver phantom is performed prior to this step. The registration is further explained in the experiments section.

The proposed haptic feedback strategy enables the user to experience viscous forces that depend on how far is the needle from the targeted tumor. 
The closer the needle is to the target, the higher the forces. A virtual wall indicates that the needle has reached the target. The forces in the control laws are represented in the axial direction of the body-fixed frame of the haptic device's handle. 

The control law of the proximity-based feedback is based on a damping coefficient $b$, the current distance to the target $d$, the offset distance $o$, and the speed of insertion $v$ as illustrated in Eq.~\ref{Eq2_proximity-based-feedback}.
\begin{equation} \label{Eq2_proximity-based-feedback}
    f_{\text{proximity}} = b v (o-d) 
\end{equation}
The controller maintains the handle's position when the user is not inserting the needle to avoid unintentional movement of the handle. 
Moreover, the device is programmed to enable one DoF movement to achieve more natural insertion.
The objective is to slow down the user during insertion to reach the target accurately.

The control law of the virtual wall is a proportional-differential (PD) position controller.
When the user tries to move the needle beyond the target position, the PD controller pushes them back to the target position. 
The control law shown in Eq.~\ref{Eq3_virtual_wall} depends on the distance $x$ from the target into the wall, the proportional gain $k_p$, the velocity of needle $v$ and the differential gain $k_d$.
\begin{equation} \label{Eq3_virtual_wall}
    f_{\text{wall}} = k_px - k_dv
\end{equation}

The source device is used by the end-user to perform needle insertion axially.
\subsection{Robotic Manipulator Control}
The robotic arm is controlled to perform surgical tasks such as needle steering and insertion. 
The desired transformation ($\mathbf{T}_{\text{desired}}$) of the end-effector  ($\Psi_{EE}$) relative to the base frame ($\Psi_{B}$) is given to the robotic manipulator using the following transformation matrix. 

\begin{equation}\label{franka_control}
    \mathbf{T}_{\text{desired}} = \mathbf{T}_{{EE}_i}^{{B}} +
\begin{bmatrix}
    &\multirow{3}{*}{\centering $\mathbf{O}_{3\times3}$}   &&\multirow{3}{*}{\centering $\mathbf{R}  \begin{bmatrix}
        d_x \\ d_y \\d_z
    \end{bmatrix}$} \\
     & & & \\
     & & & \\
    0 & 0 & 0 & 1 \\
\end{bmatrix},
\end{equation} 

where $\mathbf{T}_{{EE}_i}^{{B}}$ is the initial pose of the $\Psi_{EE}$ to $\Psi_{B}$ before motion compensation or needle insertion, and $\mathbf{O}_{3\times3}$ is a zero matrix. 
$\mathbf{R}$ is set to two different magnitudes. 
In the case of motion compensation, $\mathbf{R}$ is set to the rotation matrix of the EM frame ($\Psi_{EM}$) to $\Psi_B$ of the arm. Additionally, the estimated displacements of the AP and SI motion models update $d_y$ and $d_z$ expressed in $\Psi_{EM}$, respectively. 
During needle insertion, $\mathbf{R}$ is the rotation matrix defined from $\Psi_{EE}$ to $\Psi_{B}$ of the robotic arm. The user moves the haptic device handle to perform the displacement $d_x$ in the axial direction of the needle expressed relative to $\Psi_{EE}$~(Fig.~\ref{fig:setup}~(a)).

\subsection{Liver phantom}
The liver phantom was developed in~\cite{c22} and mimics the respiratory-induced motion in the liver. The phantom consists of a soft tissue made of silicon resin mixed with styrofoam beds that simulates the liver, and a cart that moves the liver mimicking respiration. The cart includes actuators to achieve the desired motion through pneumatic actuation. 
The liver phantom moves in two DoF simulating the SI and AP motion.
Additionally, a spherical target made of lead with a diameter of 3~mm was placed inside of the liver phantom.
The liver phantom is shown in~Fig.~1~and~Fig.~\ref{fig:setup}~(a) and (c).

\subsection{Experiments}
\subsubsection{Experimental setup}
Fig.~\ref{fig:setup}~(b) and (c) display the setup layout. 
The replica side (Fig.~\ref{fig:setup}~(c)) contains the  ARTIS pheno C-arm scan (Siemens Healthineers, Erlangen, Germany), the robotic arm 7-DoF Franka Emika (GmBH, Munich, Germany) with a needle attached to the end-effector, the EM tracker system (Northern Digital Inc., Waterloo, Canada) which is externally attached to the liver phantom.

The source side (Fig.~\ref{fig:setup}~(b)) contains the haptic device Omega.7 (Force Dimension, Nyon, Switzerland) that is operated to remotely control the robotic manipulator and provide feedback to the user based on the proximity between the needle and the target.

\subsubsection{Experimental Protocol} 
We present the experimental plan to robotically compensate for respiratory motion and perform a teleoperated insertion with haptic feedback.
\begin{itemize}
  \item \textit{Determine insertion point}: A cone beam CT 3D reconstruction of the liver phantom is obtained during a breath-hold and an insertion point is defined based on the position of the target. 
  A biopsy grid from the scan is projected on the phantom at the defined insertion site. 
  A reference needle with a length of 3~cm is inserted into the phantom following the orientation and position set by the grid. 
  The reference needle is an essential component to further align the robotic manipulator with the defined insertion point by the biopsy grid. 
    \item \textit{Manual registration}: The biopsy needle is placed at the end-effector of the robotic arm and the arm is manually moved to align the biopsy needle with the reference needle. This ensures the alignment between the robotic manipulator and the target. 
    \item \textit{Teleoperated retraction}: The robotic arm is remotely moved to a safe position away from the liver phantom in the axial direction (4~cm), and the reference needle is removed. 
    \item \textit{Motion model training}: The training of the motion model is performed using the target motion obtained from the fluoroscopic images as ground truth and the data from the EM sensor as a surrogate which was acquired simultaneously. During the training, the liver phantom simulates regular breathing and three types of breath-holds at different respiratory phases. 
    If the target was placed in a new location inside the liver phantom, a new training session was conducted.
     
      \item \textit{Motion compensation}: The breathing of the phantom starts, and the created motion model uses the EM sensor data to estimate the displacement of the tumor. 
      The estimations are given to the robotic manipulator to move the needle in the two directions of motion to keep it aligned with the target position.
    \item \textit{Teleoperated insertion}: The liver phantom stops at one of the three breath-hold positions to allow the user to perform the remote needle insertion at the source side.
    \item \textit{Validation using fluoroscopic images}: A sagittal and transverse fluoroscopic images of the liver phantom are obtained to measure the deviation between the needle tip and the target. The lateral $\epsilon_x$, top $\epsilon_y$, and depth $\epsilon_z$ errors are extracted, and the Euclidean distance was computed to measure the targeting error.  
\end{itemize}
Five insertions were performed at different breath-hold positions. Additionally, the location of the target inside the phantom did not change for the first three insertions and was moved to a different position for the remaining two. 

\section{RESULTS}
In this section, we first present the results of the respiratory motion compensation step, which include the performance of the created respiratory motion estimation models, and the robotic motion compensation. Secondly, the errors of each insertion are presented. 

\begin{figure*}[ht!]
    \centering    
    \includegraphics[scale=0.365]{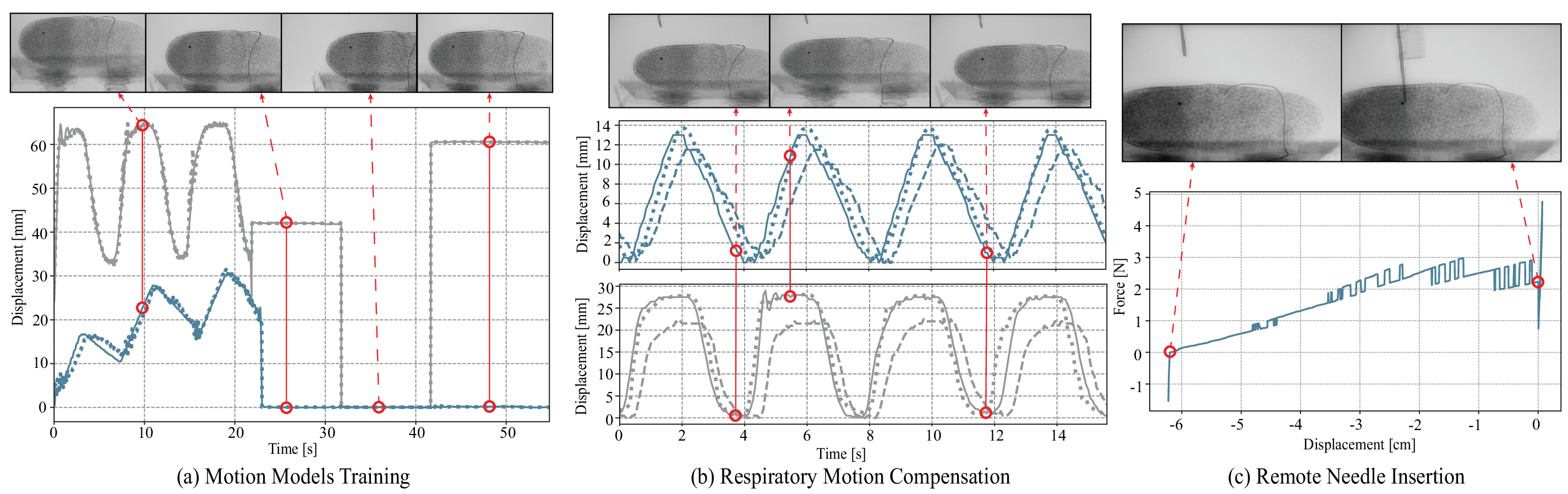}
    \vspace{-3mm}
    \caption{(a) The liver phantom simulates different types of breathing patterns during the training phase. The target location (ground truth) is extracted in the two directions of motion Anterior-Posterior and Superior-Inferior which are represented by the solid lines in blue and gray, respectively. The motion models are trained to estimate the ground truth (the dotted lines). (b) The liver phantom simulates regular breathing and the location of the target is extracted for validation (solid lines). The motion models estimate the current target position (dotted lines), and the robotic manipulator compensates for the respiratory motion by steering the needle attached to the end-effector (dashed lines). (c) The insertion is performed remotely using the haptic device. The forces felt by the user increase as the distance between the tip of the needle and the target decreases.}
    
    \label{steering}
\end{figure*}
\subsection{Respiratory Motion Compensation}
A variety of breathing patterns were simulated by the liver phantom during the training phase (Fig.~\ref{steering}~(a)).
Table~I displays the results obtained for the SI and AP motion models created for regular breathing ($M_r$) and for the different breath-hold states ($M_{b-h}$). 
The errors of the motion models were below 3~mm for the AP and SI directions of motion for the different breathing phases.    
\begin{table}[ht!]
    \small
    \centering
    \caption{Motion models results based on the mean absolute error for the train and test phases. All values are in mm.}
    \label{tab2}
    \begin{tabular}{p{2.4cm} p{1cm} p{1.8cm} p{1.8cm}}
        \toprule
        \multirow{2}{*}{Direction of Motion} & \multirow{2}{*}{} & \multicolumn{2}{c}{Motion Models} \\
        \cmidrule{3-4}
        & &\multicolumn{1}{c}{$M_r(t)$} & \multicolumn{1}{c}{$M_{b-h}(t)$} \\
        \midrule
         AP &  Train & $1.07 \pm 0.17$ & $0.15 \pm 0.21$\\
                            &  Test & $1.85\pm1.14$ & $0.07 \pm 0.03$ \\
        \midrule
         SI &  Train & $0.84 \pm 0.13$ &  $0.04 \pm 0.02$ \\
                           &  Test & $1.16\pm1.48 $ & $2.56\pm 2.65$ \\
        \bottomrule
    \end{tabular}

\end{table}

Fig.~\ref{steering}~(b) displays the performance of the motion model estimations, and the robotic needle steering during motion compensation. 
The overall error of needle steering in regular breathing was $3.00 \pm 1.81$~mm, and $7.44 \pm 4.95$~mm for the AP and SI directions of motion, respectively. 
\subsection{Needle Insertion} 
The insertion is remotely performed by the end-user using a haptic device.
The forces provided as feedback increase while the needle moves towards the tumor (Fig.~\ref{steering}~(c)).
The user feels an increase in the forces until they reach the virtual wall (5~N), which indicates they have reached the target. 
The 3D insertion error was computed from a sagittal, and transverse fluoroscopy images obtained after insertion. 
The errors are defined as the distance of the tip of the needle to the circumference of the target. 
The obtained results for the needle insertion step are displayed in Table II.  
\begin{table}
\centering
\caption{3D Insertion errors between the tip of the needle and the target. all values are in mm.}
\resizebox{\columnwidth}{!}{%
\begin{tabular}{@{}ccccc@{}}
\toprule
Insertion & $\epsilon_x$& $\epsilon_y$  & $\epsilon_z$ & Euclidean distance \\ \midrule
1 & 4.13 & 8.35 & 3.45  & 9.93 \\
2 & 1.65 & 11.58& 1.75 & 11.83\\
3 & 6.32 & 14.51 & 5.14  &  16.64\\
4 & 0.28 & 3.11 & 3.58 & 4.75 \\
5 & 0.62 & 1.19 & 0.37  & 1.39 \\ 
\midrule
Overall & $2.60\pm2.30$ & $7.75\pm5.01$ &  $2.86\pm1.64$&$8.91\pm5.35$\\ \bottomrule
\end{tabular}
}
\end{table}
\section{DISCUSSION AND CONCLUSION}
This work presents the integration of robotic respiratory motion compensation and teleoperated needle insertion with haptic feedback for X-ray-guided percutaneous procedures. 
The results of the robotic motion compensation were presented in the previous section, it was observed that the highest errors were obtained during the real-time needle steering when the liver phantom simulated regular breathing. 
Especially, in the SI direction of motion, an overall error of 7.44~mm was observed, possibly due to the higher amplitude of motion compared to the AP direction. 
Nonetheless, in this work, the insertions are only performed during a breath-hold state. In Table II, it can be observed that in the directions in which motion was compensated ($\epsilon_x$ and $\epsilon_z$), the errors are lower than 3~mm. 
Moreover, the maximum errors were obtained in insertion 3 with an Euclidean distance of 16.64~mm, and the minimum errors were observed in insertions 4 and 5 with an Euclidean distance of 4.75 and 1.39~mm.
During the first three insertions, it was observed that the liver phantom moved during the insertion due to the tool-tissue interaction. 
This error is specifically highlighted in the lateral direction of the liver phantom ($\epsilon_y$). For the final two insertions, it was observed that the liver phantom motion was limited. 

Furthermore, the radiation exposure from the physician's perspective has been eliminated since the insertion is performed remotely. 
Additionally, the provided haptic feedback during insertion is expected to reduce the need for additional check scans, thus reducing the patient's radiation exposure. 
However, further research needs to focus on minimizing the patient's radiation exposure. 
Future work should also include possible tool-tissue interactions that consider needle deflections during insertion. 
These aspects should be incorporated into motion models to accurately estimate the position of the target. 
Also, experimental validation could be extended by performing animal or cadaveric studies and by including interventional radiologists as end-users. 

It can be concluded from this study that respiratory motion models can compensate for breathing changes to minimize the chances of misdiagnosis or incomplete treatment in percutaneous procedures. 
Additionally, teleoperated insertion with haptic feedback provides the guidance needed to complete an insertion while potentially reducing radiation exposure in X-ray-guided interventions.
\section*{ACKNOWLEDGMENT}
The work in this paper was supported by ITEA under
the 20044 ASSIST project. The authors would like to thank
Remco Liefers for his contribution during the experiments,
Yoeko Xavier Mak for his contribution on the preparation of
the experiments, Girindra Wardhana for his contribution in
providing graphics and Youssef Aboudorra for reviewing the
manuscript.

\end{document}